\documentclass[reprint,amsmath,amssymb,twocolumn,showkeys, superscriptaddress,floatfix,showkeys]{revtex4-2}
\usepackage{graphicx}% Include figure files
\usepackage{dcolumn}% Align table columns on decimal point
\usepackage{bm}% bold math
\usepackage[colorlinks = true, allcolors = blue]{hyperref}% add hypertext capabilities
\usepackage{verbatim}
\usepackage{soul}
\usepackage{float}
\usepackage{xcolor}
\usepackage{multirow}

\begin{document}

\title{Analytical Retrieval of Material Parameters in Monolayer Transition-Metal Dichalcogenides Based on a Solvable Exciton Model}

\author{Duy-Nhat Ly}
\email{nhatld@hcmue.edu.vn}
\affiliation{Computational Physics Key Laboratory K002, Department of Physics, Ho Chi Minh City University of Education, Ho Chi Minh City 72759, Vietnam}

%\author{Quoc-Huy Nguyen}
%\affiliation{Computational Physics Key Laboratory K002, Department of Physics, Ho Chi Minh City University of Education, Ho Chi Minh City 72759, Vietnam}

\author{Thanh-Son Nguyen}
\affiliation{Faculty of Fundamental Sciences, University of Architecture Ho Chi Minh City, Ho Chi Minh City 722700, Vietnam}

\author{Ngoc-Tram D. Hoang}
\affiliation{Computational Physics Key Laboratory K002, Department of Physics, Ho Chi Minh City University of Education, Ho Chi Minh City 72759, Vietnam}

\author{Van-Hoang Le}
\email{hoanglv@hcmue.edu.vn}
\affiliation{Computational Physics Key Laboratory K002, Department of Physics, Ho Chi Minh City University of Education, Ho Chi Minh City 72759, Vietnam}

\date{\today}% It is always \today, today,

\begin{abstract}

We develop an analytical procedure to retrieve fundamental material parameters of monolayer transition-metal dichalcogenides from optical and magneto-optical exciton spectra, based on the solvable modified Kratzer model. The proposed retrieval procedure naturally consists of two complementary stages. In the first stage, explicit inversion formulas determine the quasiparticle bandgap, effective screening parameter, and energy scaling factor directly from the experimentally measured energies of the three lowest excitonic states, from which the screening length is subsequently obtained. In the second stage, an analytical expression for the magnetic-field dependence of the exciton energies independently yields the reduced exciton mass, from which the surrounding dielectric constant is then calculated. Once the complete set of material parameters has been retrieved, the framework analytically predicts the diamagnetic coefficients, exciton radii, and complete magnetoexciton spectra without introducing additional fitting parameters or matrix diagonalization. The method is applied to a broad range of experimental samples for WSe$_2$, WS$_2$, MoS$_2$, MoSe$_2$, and MoTe$_2$ monolayers embedded in different dielectric environments. The retrieved material parameters are in good agreement with independent experimental measurements and previous Rytova--Keldysh (RK) calculations, while the predicted excitonic properties accurately reproduce available magneto-optical observations. The proposed analytical theory provides an efficient, physically transparent alternative to conventional numerical fitting procedures and offers an effective tool for the rapid characterization of two-dimensional semiconductors via excitonic spectroscopy.

\end{abstract}

\maketitle

%========================================================
\section{Introduction}
%========================================================

Monolayer transition-metal dichalcogenides (TMDCs) have appeared as an important class of two-dimensional semiconductors owing to their direct bandgaps, strong Coulomb interactions, and pronounced excitonic effects arising from reduced dielectric screening and quantum confinement \cite{Wang2018,Chen2023,Biswas2023,Champagne2023,Gillespie2024,Pico2025,Eunice2024,Huang2026}. Their exciton binding energies, typically in hundreds of meV range, enable stable excitonic states even at room temperature, making these materials attractive for both fundamental studies and optoelectronic applications \cite{berkelbach2013,chernikov2014,HePRL2014,Hill2015,Stier2016,Chennano2019,Liu2021,Kezerashvili2021,nhat2023Apr,Zhu2024,Li2025}. Since exciton energies are governed by the underlying electronic structure and dielectric environment, optical and magneto-optical spectroscopy provide powerful tools for probing the fundamental material parameters of TMDC monolayers, including the quasiparticle bandgap, screening length, dielectric constant of the surrounding environment, and reduced exciton mass \cite{Stier2018,Liu2019,NGUYEN2019,NAT2019,Chennano2019,Hsu2DMater2019,Nhat2023May,Hu2024,takahashi2024,Seksaria2024}.

Among the available theoretical descriptions, the Rytova-Keldysh (RK) potential \cite{Rytova1967, keldysh1979} has become the standard model for describing the screened electron-hole interaction in two-dimensional semiconductors \cite{berkelbach2013,chernikov2014}. Combined with accurate numerical solutions of the excitonic Schrödinger equation, the RK model successfully reproduces a broad range of experimental observations. Moreover, the inverse problem of extracting material parameters from experimentally measured exciton spectra has been solved, typically via iterative fitting~\cite{Stier2018,Liu2019,NGUYEN2019,NAT2019,Nhat2023May}. Such numerical procedures vary several correlated parameters simultaneously until the best fit to the experiment is achieved; therefore, they are computationally expensive and often conceal the physical role of individual parameters, particularly for comparing different dielectric environments or combining optical and magneto-optical measurements.

To overcome these limitations, we previously developed an analytical retrieval method based on an approximate analytical solution of the RK model for extracting the material parameters of hBN-encapsulated TMDC monolayers directly from experimental exciton spectra \cite{Dinh2025,Nhat2025}. That work demonstrated that analytical inversion can provide an efficient alternative to conventional numerical fitting. However, because the analytical approximation was derived for a specific range of dielectric screening, its applicability is restricted to a relatively narrow class of dielectric environments.

An attractive alternative for describing the exciton spectrum of monolayer TMDCs was proposed by Molas et al. \cite{Molas2019} using the solvable modified Kratzer potential~\cite{Kratzer1920}. Although this model is considerably simpler than the RK one, it reproduces the low-lying exciton spectrum with remarkable accuracy over a broad range of experimentally relevant dielectric environments. Its analytical solvability yields explicit expressions for both exciton energies and wave functions, making it particularly well-suited to developing analytical inversion methods and establishing direct relationships between experimentally measured spectra and the underlying material parameters.

In the present work, we develop a complete analytical framework for retrieving fundamental material parameters of monolayer TMDCs using the solvable modified Kratzer model~\cite{Kratzer1920,Molas2019}. The retrieval procedure naturally consists of two complementary stages. In the first stage, explicit inversion formulas determine the quasiparticle bandgap energy, effective screening parameter, energy scaling factor, and, consequently, the screening length directly from the experimentally measured energies of the three lowest excitonic states. In the second stage, the reduced exciton mass is determined independently from the magnetic-field dependence of the exciton energies through an analytical magnetoexciton expression, from which the surrounding dielectric constant is subsequently obtained. Once the complete set of material parameters has been retrieved, the framework further predicts the diamagnetic coefficients, exciton radii,  and magnetoexciton spectra analytically, without introducing additional fitting parameters or numerical diagonalization. The complete retrieval workflow is summarized schematically in Fig.~\ref{fig2}.

Compared to our previous analytical theory based on the Rytova-Keldysh potential, this proposed framework extends the applicability of material parameter retrieval to a broader range of dielectric environments. At the same time, it still keeps computational efficiency and physical transparency. The retrieved parameters are validated for representative TMDC monolayers embedded in various dielectric environments by comparison with available experimental measurements and our previous RK-based retrieval results. Besides providing an efficient analytical alternative to conventional numerical fitting procedures, the present framework establishes a quantitative bridge between the analytically solvable modified Kratzer model and the microscopic RK description, thereby offering a practical tool for rapid characterization of two-dimensional semiconductors from optical and magneto-optical spectroscopy.

The rest of this paper is organized as follows. Section~\ref{sec:2} presents the analytical retrieval framework. It derives inversion formulas to determine the zero-field material parameters, develops an analytical description of magnetoexciton energies to retrieve the reduced exciton mass, and summarizes the complete two-stage retrieval workflow. Section~\ref{sec:3} applies the proposed methodology to representative TMDC monolayers in different dielectric environments, where the retrieved material parameters, analytically predicted excitonic properties, and comparisons with experimental measurements and numerical calculations are presented and discussed. Finally, Sec.~\ref{sec:4} summarizes the principal conclusions of this work.

\section{Analytical Procedure for Retrieving Material Parameters from Exciton Energies }
\label{sec:2}

\subsection{The solvable exciton spectrum in the modified Kratzer model}
\label{subsec:2a}

Within the modified Kratzer model introduced by Molas \textit{et al.} \cite{Molas2019}, the excitonic binding energies of the $ns$ states in monolayer TMDCs are described by the analytical expression:
\begin{equation}\label{eq1}
\varepsilon_{n}=-\frac{\eta }
{\left(n-{1}/{2}+\xi\right)^2}\text{Ry}.
\end{equation}
Here, $\text{Ry}$ is the Rydberg unit of energy; the dimensionless parameter 
\begin{equation}\label{eq2}
\eta=\frac{\mu}{\kappa^2}
\end{equation} 
is the rescaling factor of the Rydberg energy, 
where $\mu$ is the reduced exciton mass in units of the electron mass $m_e$, and $\kappa$ is the dielectric constant of the surrounding medium.  

In Eq.~\eqref{eq1}, $\xi$ is the dimensionless parameter capturing the screening effect, which we call the effective screening parameter. In the work \cite{Molas2019}, this parameter relates to the screening length $r_0$ by the formula 
\begin{equation}\label{eq3}
\xi=g \sqrt{\eta\frac{r_0}{a_0}},
\end{equation}
where $a_0$ is the Bohr radius; $g$ is the coefficient in the modified Kratzer potential, which is typically a material-specific parameter. 
Molas \textit{et al.} used $g^2=0.21$ for WSe$_2$ encapsulated in hBN. 
We examined the applicability of Eqs.~\eqref{eq1} and \eqref{eq3} to a large number of TMDC monolayers embedded in different dielectric environments. We found that the experimental exciton energies are well reproduced by taking $g^2$ in the narrow interval $0.195 - 0.216$, i.e., $0.205 (\pm 5\%)$. Therefore, throughout this work, we adopt the universal value $g^2=0.205$, which provides a good balance between universality and accuracy.

\subsection{Inversion formulas for material parameters from zero-field exciton spectrum}
\label{subsec:2b}

In the experimental photoabsorption spectrum from the TMDC monolayers, the measured exciton energies are
\begin{equation}\label{eq4}
E_n=\varepsilon_n(\eta, \xi)+E_g,
\end{equation}
where $E_g$ is the quasiparticle bandgap energy, while $\varepsilon_n (\eta, \xi)$ is the excitonic binding energy estimated by Molas's formula \eqref{eq1}. The lowest three $ns$ exciton energies are particularly sensitive to dielectric screening \cite{chernikov2014} and therefore contain sufficient information for determining the unknown material parameters. This observation is consistent with Eq.~\eqref{eq1}, in which the exciton energies exhibit a strong dependence on the parameters $\eta$ and $\xi$ for low principal quantum number $n$. Suppose that only the experimentally measured $s$-state exciton energies $E_1$, $E_2$, $E_3$ are available, while the material parameters remain unknown. Equation~\eqref{eq4} then provides a direct route for retrieving the material parameters $E_g$, $\eta$, and $\xi$. Since there are three unknown parameters, three measured exciton energies are sufficient to formulate the inverse problem.

\begin{figure}[tbp!]
\center
\includegraphics[width=0.95 \columnwidth]{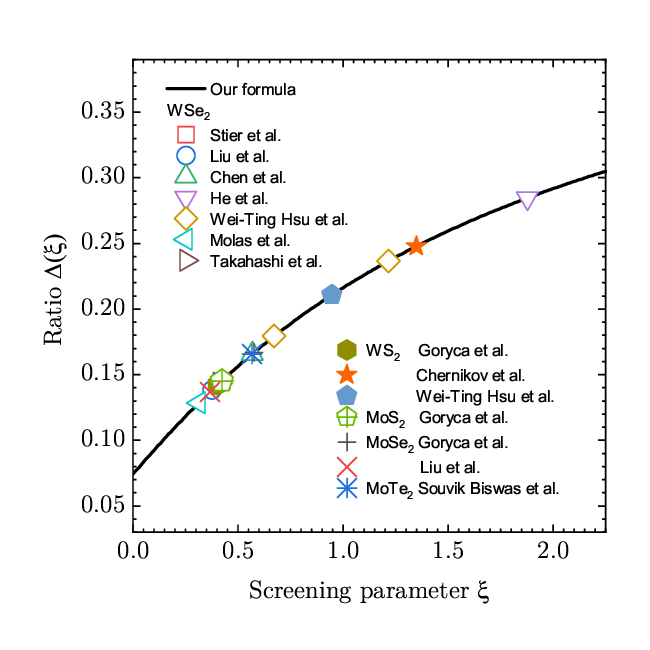}
\caption{Ratio $\Delta(\xi)$ as a function of the screening parameter $\xi$ (solid line), compared with the experimental values $\Delta_{\mathrm{exp}}$ for different TMDC monolayers (symbols).}
\label{fig1}
\end{figure}

First, we define an experimental value
\begin{equation}\label{eq5}
\Delta_{exp}=\frac{E_{3}-E_{2}}
{E_{3}-E_{1}},
\end{equation}
whose value lies in the interval $0<\Delta_{exp} < 1$ since $E_1 < E_2 < E_3$. From Eq.~\eqref{eq4}, we can see that the difference between excitonic energies equals the difference of the corresponding binding exciton energies: $E_j-E_k =\varepsilon_j-\varepsilon_k $. Therefore, Eq.~\eqref{eq5} leads to an equation
\begin{equation}\label{eq6}
\Delta_{exp}=\frac{\varepsilon_{3}-\varepsilon_{2}}
{\varepsilon_{3}-\varepsilon_{1}}=\frac{1}{2}-\frac{3}{2(2\xi+3)}+\frac{2}{(2\xi+3)^3}.
\end{equation}
Equation \eqref{eq6} is the key equation in the present analytical retrieval procedure, as it determines the effective screening parameter independently of both the band gap and the energy scaling factor.
Figure~\ref{fig1} plots the ratio $\Delta (\xi)={(\varepsilon_{3}-\varepsilon_{2})}/{(\varepsilon_{3}-\varepsilon_{1})}$ as a function of $\xi$ and presents various experimental values of $\Delta_{exp}$. We can see that the function $\Delta(\xi)$ is monotonic and that Eq.~\eqref{eq6} has a single solution, whose analytical form is derived below.

% Figure 1

By introducing the variable $X=1/(2\xi+3)$, we can transform Eq.~\eqref{eq6} into a cubic equation with three analytic solutions. However, to simplify the cubic equation and obtain a more compact analytical expression, we introduce the auxiliary variable $\lambda=(2\xi+1)\delta/(2\xi+3)$ with $\delta=\sqrt{\Delta_{exp}}$. Equation~\eqref{eq6} thereby becomes the cubic equation: $3\lambda-4\lambda^3=\delta$. Although this cubic equation possesses three real roots, only one satisfies the physical condition $\xi>0$. Solving this equation by the trigonometric method, we obtain 
\begin{equation}\label{eq7}
\lambda=\cos\left(\frac{1}{3}\arcsin\delta+\frac{\pi}{6}\right)\;\;\text{with} \;\;\delta=\sqrt{\Delta_{exp}}
\end{equation}
and consequently
\begin{eqnarray}\label{eq8}
\xi=\dfrac{3\delta-\lambda}{2(\lambda-\delta)},
\end{eqnarray}
where the physical solution exists only in the interval
\begin{equation}\label{eq9}
\sqrt{\frac{2}{27}}< \delta < \frac{1}{\sqrt{2}}.
\end{equation}

Now, knowing the effective screening parameter $\xi$, we can determine the bandgap energy $E_g$ and the scaling factor $\eta$ from two equations from Eq.~\eqref{eq4} for $E_1$ and $E_2$. Using two quantities $\delta$ and $\lambda$ instead of $\xi$ in Eq.~\eqref{eq8}, we have 
\begin{eqnarray}\label{eq10}
\eta&=&\frac{\lambda^2 \delta^2}{(\lambda-\delta)^2(\lambda^2-\delta^2)}\frac{E_2-E_1}{\text{Ry}},\\
E_g&=&\frac{\lambda^2 E_2-\delta^2 E_1}{\lambda^2-\delta^2}.\label{eq11}
\end{eqnarray}

Consequently, the three measured low-lying exciton energies determine the effective screening parameter $\xi$, the energy scaling factor $\eta$, and the quasiparticle band gap $E_g$. The screening length $r_0$ then follows directly from Eq.~\eqref{eq3}, yielding the formula
\begin{equation}\label{eq12}
r_0=\frac{1}{g^2\eta}\xi^2 a_0,
\end{equation}
where $g^2=0.205$, and $a_0$ is the Bohr radius. 
The reduced exciton mass $\mu$ is determined independently from magnetoexciton energies, as described in Subsec.~\ref{subsec:2c}. Once $\mu$ is known, the dielectric constant $\kappa$ follows directly from Eq.~\eqref{eq2}, which leads to
\begin{equation}\label{eq13}
\kappa=\sqrt{\frac{\mu}{\eta}}.
\end{equation}

\subsection{Analytical method for determining the reduced exciton mass from magnetoexciton energies}
\label{subsec:2c}

This section completes the analytical retrieval procedure by determining the remaining unknown material parameter, namely the reduced exciton mass.

The reduced exciton mass $\mu$ is determined from the magnetic-field dependence of the exciton energies. In our previous work \cite{Nhat2025}, we derived an analytical interpolation formula for the magnetoexciton spectrum based on the Rytova–Keldysh potential. Here, we adopt the same interpolation structure, but replace the zero-field exciton energies and diamagnetic coefficients by analytical expressions derived from the solvable modified Kratzer model. The magnetoexciton energies for $ns$ states are therefore written as
\begin{equation}\label{eq14}
E_n (B)=E_g+\varepsilon_n (\eta,\xi)+\frac{\sigma_n B^2}{1+\sigma_n\beta_n^{-1} B},
\end{equation}
where $\sigma_n$ is the diamagnetic coefficient and $\beta_n$ is the strong-field slope parameter that ensures the correct high-field asymptotic behavior.

Within first-order perturbation theory, the diamagnetic coefficient of the $ns$ state can be calculated by the formula
\begin{equation}\label{eq15}
\sigma_n=\frac{e^2}{4\mu^*}   r_n^2 =\frac{1}{4\mu}   \langle r^2 \rangle_n \frac{{\mu_B}^2}{\text{Ry}},
\end{equation}
where $\mu^*=\mu m_e$ is the reduced exciton mass. Meanwhile, $r_n^2= \langle r^2 \rangle_n $ denotes the mean-square exciton radius (in the units of $a_0^2$) defined as
\begin{eqnarray}\label{eq16}
\langle r^{2}\rangle_{n}
= \int\limits_{0}^{+\infty} \mathcal{R}_{n}^{2}(r)\, r^{3}\,dr 
\end{eqnarray}
with the field-free $s$-state radial wave function obtained within the solvable modified Kratzer model \cite{Molas2019} as
\begin{eqnarray}\label{eq17}
\mathcal{R}_{n}(r)= \mathcal{A}_{n}\,\omega^{\xi+1} \,\mathrm{e}^{-\omega r/2}\,r^{\xi}\,
\mathcal{L}_{n}^{2\xi}(\omega r).
\end{eqnarray}
Here, $\mathcal{A}_{n} = \sqrt{ n!/(n+2\xi)!/(2n+2\xi+1) }$ is the normalization constant, $\mathcal{L}_{n}^{2\xi}(x)$ denotes the generalized Laguerre polynomials, and $\omega = 4/(2n+2\xi+1)\kappa$, see Appendix~\ref{AppA}. Substituting the radial function \eqref{eq17} into Eq.~\eqref{eq16} yields the analytical result 
\begin{eqnarray}\label{eq18}
\sigma_{n}= \frac{ \alpha_n(\xi) }{8 \eta\mu^{2}}  \frac{{\mu_B}^2}{\text{Ry}},
\end{eqnarray}
where $\mu_B=e\hbar/2m_e$ is the Bohr magneton, $\alpha_n(\xi)$ is given by 
\begin{eqnarray}\label{eq19}
\alpha_n (\xi)=\left(  n+\xi-{1}/{2}  \right)^4 \left[ 5-\frac{3\xi^2-7/4}{\left(  n+\xi-1/2   \right)^2}   \right] .
\end{eqnarray}

On the other hand, the strong-field slope parameter is defined by
\begin{eqnarray}\label{eq20}
\beta_{n} =\dfrac{1}{2\mu}  (2nc - 1) \mu_B,
\end{eqnarray}
so that the magnetoexciton energy described by Eq.~\eqref{eq14} correctly reproduces both weak- and strong-field limits. 
The factor $c $ controls the crossover to the high-field Landau-level limit. Although $c=1$ in the asymptotic limit, comparison with exact numerical calculations shows that $c=0.905$ provides excellent agreement for magnetic fields up to approximately 90 T. This calibrated value is adopted throughout the present work.

The analytical magnetoexciton formula \eqref{eq14} provides a direct approach for determining the reduced exciton mass $\mu$ from experimental magnetoexciton energies. For convenience, we rewrite the magnetic contribution to the exciton energies $\Delta E_n (B)={\sigma_n B^2}/{(1+\sigma_n\beta_n^{-1} B)}$ in the form
\begin{eqnarray}\label{eq21}
\Delta E_n (\gamma)=
\frac{1}{8\mu}\frac{\alpha_n \gamma^2}{\eta\mu+\dfrac{\alpha_n\gamma}{4(2nc-1)}} {\text{Ry}},
\end{eqnarray}
where $\gamma$ is the dimensionless magnetic-field strength, measured in units of $B_0={\text{Ry}}/\mu_B$. 

For each pair of measured exciton energies at magnetic fields $B_1$ and $B_2$ (in the same exciton state), we define the dimensionless experimental quantity 
\begin{equation}\label{eq22}
\delta_{mag}^{exp}=\frac{E_n(B_1)-E_n(B_2)}{\alpha_n \mu_B(B_1-B_2)}.
\end{equation}
This quantity can be calculated directly from the measured magnetoexciton energies. On the other hand, the same quantity can also be evaluated theoretically by using the analytical magnetoexciton energies. Since
$E_n (B_1)-E_n (B_2)=\Delta E_n (B_1)-\Delta E_n (B_2)$, the theoretical quantity $\delta_{mag}^{theo}$ is obtained by substituting Eq.~\eqref{eq21} for $\Delta E_n$. Equating the theoretical and experimental values, 
\begin{equation}\label{eq23}
\delta_{mag}^{theo} (\mu)=\delta_{mag}^{exp}, 
\end{equation}
yields a cubic equation for the reduced exciton mass,
\begin{equation}\label{eq24}
\mu^3+A\mu^2+B\mu+C=0
\end{equation}
with
\begin{eqnarray}\label{eq25}
A&=&\frac{\alpha_n(\gamma_1+\gamma_2)}{4\eta(2nc-1)},\nonumber\\
B&=&\frac{\alpha_n^2\gamma_1\gamma_2}{16\eta^2(2nc-1)^2}-\frac{\gamma_1+\gamma_2}{8\eta\delta_{mag}^{exp}},\\
C&=&-\frac{\alpha_n\gamma_1\gamma_2}{32\eta^2(2nc-1)\delta_{mag}^{exp}}.\nonumber
\end{eqnarray}
Solving Eq.~\eqref{eq24} for each pair of magnetic-field values yields one estimate of the reduced exciton mass. Averaging the values obtained from all available experimental data provides the final value of $\mu$.

Alternatively,  instead of solving Eq.~\eqref{eq24}, the reduced exciton mass can be determined by fitting the analytical expression \eqref{eq21} directly to the measured magnetoexciton energies. However, unlike our previous implementation \cite{Nhat2023May}, where each trial value of $\mu$ required numerically solving the Schr{\"o}dinger equation, the present approach is entirely analytical, making the fitting procedure significantly faster and more convenient.

\subsection{Two-stage analytical retrieval workflow}
\label{subsec:2d}

\begin{figure}[htbp!]
\center
\includegraphics[width=1.05 \columnwidth]{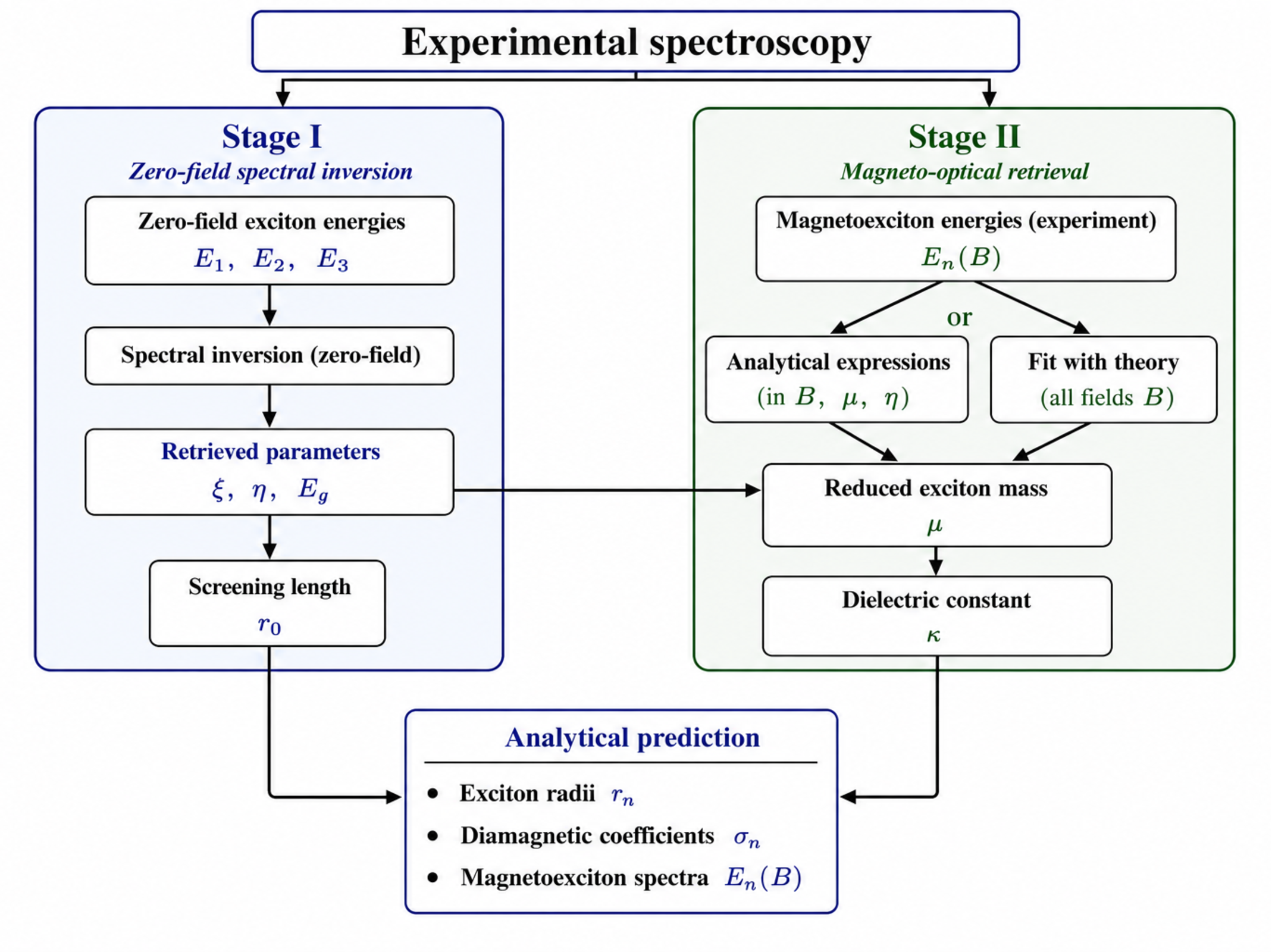}
\caption{Schematic flowchart of the analytical retrieval framework. The workflow retrieves the fundamental material parameters (bandgap energy $E_g$, screening length $r_0$, reduced exciton mass $\mu$, and surrounding dielectric constant $\kappa$) from experimental exciton and magnetoexciton energies and predicts the derived excitonic properties, including the diamagnetic coefficients $\sigma_n$, exciton radii $r_n$, and magnetoexciton spectra $E_n(B)$.}
\label{fig2}
\end{figure}

The complete analytical retrieval workflow developed in the present work is summarized schematically in Fig.~\ref{fig2}. It consists of two complementary retrieval stages, namely zero-field spectral inversion and magneto-optical determination of the reduced exciton mass, followed by an analytical prediction stage.

In the first stage, the experimentally measured zero-field exciton energies $(E_{1}, E_{2}, E_{3})$ are used as input. Applying Eqs.~\eqref{eq8}--\eqref{eq11} determines the effective screening parameter $\xi$, the energy scaling factor $\eta$, and the bandgap energy $E_g$. The screening length $r_0$ then follows directly from Eq.~\eqref{eq12}. 

In the second stage, the experimentally measured magnetoexciton energies at different magnetic fields serve as the input. The quantity $\delta_{{mag}}^{{exp}}$, defined by Eq.~\eqref{eq22}, is constructed from each pair of magnetic-field measurements and compared with its analytical counterpart calculated from Eq.~\eqref{eq21}. Their equality leads to the cubic equation~\eqref{eq24}, whose physical solution yields the reduced exciton mass. Repeating the procedure for all available pairs of magnetic-field measurements and averaging the resulting values provides the final estimate of $\mu$. The dielectric constant $\kappa$ of the surrounding environment is then obtained directly from Eq.~\eqref{eq13}.

Once the complete set of material parameters $E_g, \xi, \eta, \mu$ (or $E_g, r_0, \kappa, \mu$) has been determined, the analytical model directly yields the derived excitonic properties. In particular, the diamagnetic coefficients $\sigma_n$, exciton radii $r_n$, and magnetoexciton spectra $E_n (B)$ are obtained analytically from Eqs.~\eqref{eq15}--\eqref{eq21} without any additional fitting parameters or numerical diagonalization. Thus, the present framework not only retrieves the fundamental material parameters but also analytically predicts the principal excitonic properties.

% Figure 2

This two-step retrieval procedure offers several practical advantages. First, the zero-field exciton spectrum determines the parameters $(E_g,\xi,\eta)$ independently of the magnetic-field measurements, while the magneto-optical spectra are used only to retrieve the reduced exciton mass $\mu$. Therefore, even when limited magnetic-field data are available, the complete set of material parameters can still be determined. Furthermore, using multiple excitonic states significantly improves the reliability of exciton mass retrieval. It can be explained by the fact that diamagnetic coefficients increase rapidly with the principal quantum number as shown in Eq.~\eqref{eq19}, making them considerably more sensitive to the reduced exciton mass. Consequently, simultaneous retrieval of $\mu$ from different excitonic states enhances overall accuracy and also provides an important internal consistency check of the analytical framework.

%========================================================
\section{Results and Discussion}
\label{sec:3}
%========================================================

\subsection{Retrieval of material parameters from experimental exciton spectra}\label{subsec:3a}

In this subsection, the analytical retrieval framework developed in Sec.~\ref{sec:2} is applied to monolayer transition-metal dichalcogenides (TMDCs) embedded in different dielectric environments in order to determine their fundamental material parameters from measured exciton and magnetoexciton spectra. While our previous work \cite{Nhat2025} focused on hBN-encapsulated TMDC monolayers, the present study extends the analysis to a broad range of TMDC materials and dielectric environments using the optical and magneto-optical spectra reported in Refs.~\cite{chernikov2014, HePRL2014, Stier2018, Hsu2DMater2019, Molas2019, Liu2019, Chennano2019, NAT2019, Biswas2023, Champagne2023,takahashi2024}. These experimental and theoretical data provide the basis for validating the proposed retrieval framework and its analytical predictions.

Tables~\ref{tab1} and \ref{tab2} summarize the complete analytical retrieval of the material parameters (the band gap, screening length, reduced exciton mass, and surrounding dielectric constant) for the investigated TMDC monolayers. Table~\ref{tab1} focuses  on WSe$_2$ monolayer embedded in hBN and on some cases where the monolayer lays on other substrates (SiO$_2$, PDMS, and Sapphire). Tabe~\ref{tab2} considers WS$_2$ and other Mo-based TMDC monolayers embedded in hBN. Other environments are also investigated in two cases for WS$_2$ on SiO$_2$ and Sapphire substrates. 

For all datasets, the values $E_g$ and $r_0$  are obtained directly from three measured exciton energies, $E_{1s}$, $E_{2s}$, and $E_{3s}$, by the procedure in the first stage as described in Subsec.~\ref{subsec:2d}. 
The higher excitonic energies $E_{4s}$ and $E_{5s}$ are then predicted analytically and included for comparison. 

The determination of the reduced exciton mass depends on the available experimental information. For experiments in which magneto-optical spectra are available, the reduced exciton mass $\mu$ is retrieved independently from the analytical magnetoexciton theory developed in Sec.~\ref{sec:2}, and the effective dielectric constant $\kappa$ is subsequently obtained by Eq.~\ref{eq13}. For experiments without magnetic-field measurements, the surrounding dielectric constant $\kappa$ is fixed according to the known dielectric environment or independently established values reported in the literature, while the reduced exciton mass is then determined from Eq.~\eqref{eq2}. 

In this way, Tables~\ref{tab1} and \ref{tab2} present not only the experimentally measured and analytically predicted exciton energies, but also the complete set of retrieved material parameters $(E_g,r_0,\mu,\kappa)$ together with the corresponding experimental or theoretical benchmark values available in the literature. The detailled discussion are provided in follows.

% Table  I
\begin{table*}[htbp]
\caption{\label{tab1} Experimental and predicted $s$-state excitonic energies, together with the retrieved material parameters $E_g$, $r_0$, $\mu$, and $\kappa$, for WSe$_2$ monolayers in different dielectric environments, compared with the corresponding experimental and theoretical values reported in the literature.}
\begin{ruledtabular}
\begin{tabular}{llccc cc cccc}
Reference &Environment &\multicolumn{5}{c}{Experimental and predicted energies (eV)} &
\multicolumn{4}{c}{Retrieved parameters} \\
\cline{3-7}\cline{8-11}
\noalign{\vskip2pt}
& &$E_{1}$ &$E_{2}$ &$E_{3}$ &$E_{4}$ &$E_{5}$ &$E_g$ (eV) &$r_0$ (nm) &$\mu$ &$\kappa$ \\
\hline
\noalign{\vskip3pt}
Stier et al.~\cite{Stier2018}& hBN/hBN & & & & & & & & & \\
\quad -- Exp. &     & 1.723\,\,\; & 1.853\,\,\;& 1.875\,\,\; &-- & --&--  & -- &--  & -- \\
\quad -- Theory & & 1.729\,\,\; & 1.853\,\,\; & 1.874 \,\,&-- &-- & 1.890        &4.50    &$0.20\pm 0.01 $   &4.50 \\
Retrieval~\cite{Nhat2025} & & 1.723\,\,\; & 1.853\,\,\; & 1.875\,\,\; & -- & -- & 1.892 & 4.35 & 0.19\,\,\; & 4.25 \\
\textbf{This work}& & 1.723\,\,\; & 1.853\,\,\; & 1.875\,\,\; & 1.883\,\,\;  & 1.886\,\,\; & 1.892 & 4.34 & 0.20 & 4.36 \\[3pt]

Liu et al.~\cite{Liu2019}
& hBN/hBN & & & & & & & & & \\
\quad -- Exp.  &     & 1.7120\,\,\; & 1.8427\,\,\; & 1.8642\,\,\; & 1.8732\,\,\; & --&-- & -- & -- & -- \\
\quad -- Theory & & 1.7077     & 1.8397      & 1.8634     & 1.8721        & --& 1.884     & 5.00     &0.20\,\,\;            &3.97\\
Retrieval~\cite{Nhat2025} & & 1.712\,\,\; & 1.843\,\,\; & 1.864\,\,\; & -- & -- & 1.879 & 3.89 & 0.185 & 4.47 \\
\textbf{This work} & & 1.7120 & 1.8427 & 1.8642& 1.8714 & 1.8747& 1.880 & 4.09 & 0.195 & 4.43 \\[3pt]

Takahashi et al.~\cite{takahashi2024}& hBN/hBN & & & & & & & & & \\
\quad -- Exp.& & 1.6982 & 1.8273 & 1.8489 &-- &-- & -- & -- & -- & -- \\
\quad -- Theory& & --& -- & --&-- &-- & 1.862 & $3.45\pm 0.82$ & 0.20\,\,\; & $5.01\pm 0.66$ \\
Retrieval~\cite{Nhat2025} & & 1.6982 & 1.8273 & 1.8489 & -- & -- & 1.865 & 4.29 & 0.20\,\,\; & 4.33 \\
\textbf{This work}& & 1.6982 & 1.8273 & 1.8489 & 1.8562 & 1.8595 & 1.865 & 4.28 & 0.21\,\,\; & 4.5 \\[3pt]

Chen et al.~\cite{Chennano2019}& hBN/hBN & & & & & & & & & \\
\quad -- Exp.&     & 1.727\,\,\; & 1.858\,\,\; & 1.884\,\,\; &1.918 &-- & -- & -- &-- &--  \\
\quad -- Theory& & 1.731\,\,\; & 1.859\,\,\; & 1.882\,\,\; & --&-- & 1.900      & 4.51  & 0.22\,\,\;       &4.50 \\
Retrieval~\cite{Nhat2025} & & 1.727\,\,\; & 1.858\,\,\; & 1.884\,\,\; & -- & -- & 1.907& 5.52 & 0.19\,\,\; & 3.35 \\
\textbf{This work}& & 1.727\,\,\; & 1.858\,\,\; & 1.884\,\,\; & 1.893 \,\,\; & 1.898\,\,\;  & 1.906 & 5.56 & 0.20\,\,\; & 3.65 \\[3pt]

Molas et al.~\cite{Molas2019}& hBN/hBN & & & & & & & & & \\
\quad -- Exp.& & 1.711\,\,\; & 1.840\,\,\; & 1.859\,\,\; & --&-- &  & &  &  \\
\quad -- Theory& & 1.7052 & 1.8346 & 1.8564 & 1.8638 & 1.8672 &1.873 & 4.50& 0.20\,\,\; & 4.50\\
Retrieval~\cite{Nhat2025}& & 1.711\,\,\; & 1.840 \,\,\;& 1.859\,\,\; & --  & -- & 1.873 & 4.26 & 0.19\,\,\; & 4.35 \\
\textbf{This work}& & 1.711\,\,\; & 1.840 \,\,\;& 1.859\,\,\; & 1.865\,\,\;  & 1.868 \,\,\; & 1.873 & 3.23 & 0.16\,\,\; & 4.50 \\[3pt]

He et al.~\cite{HePRL2014}& Vacuum/SiO$_2$ & & & & & & & & & \\
\quad -- Exp.& & 1.651\,\,\; & 1.817\,\,\; & 1.883\,\,\; & 1.914\,\,\; & 1.942\,\,\; & 2.020 & --& --& 1.55 \\
\textbf{This work}& & 1.651\,\,\; & 1.817\,\,\; & 1.883\,\,\; & 1.916\,\,\;  & 1.934\,\,\;  & 1.980 & 6.65 & 0.33 & 1.55 \\[3pt]

Hsu et al.~\cite{Hsu2DMater2019}& Vacuum/PDMS & & & & & & & & & \\
\quad -- Exp.& & 1.659\,\,\; & 1.882\,\,\; & 1.951\,\,\; & --&-- & 2.030 & 4.44 & 0.17\,\,\; & 1.49 \\
\textbf{This work}& & 1.659\,\,\; & 1.882\,\,\; & 1.951\,\,\; & 1.981\,\,\; & 1.997 & 2.030 & 4.75 & 0.18 & 1.49 \\[3pt]

Hsu et al.~\cite{Hsu2DMater2019}& Vacuum/Sapphire & & & & & & & & & \\
\quad -- Exp.& & 1.659\,\,\; & 1.854\,\,\; & 1.907\,\,\; & --& --& 1.960 & 4.55 & 0.17\,\,\; & 2.07 \\
\textbf{This work}& & 1.659\,\,\; & 1.854\,\,\; & 1.907\,\,\; & 1.929 & 1.940\,\,\; & 1.961 & 5.08 & 0.21\,\,\; & 2.07 \\

\end{tabular}
\end{ruledtabular}
\end{table*}

% Table II
\begin{table*}[htbp]
\caption{\label{tab2}
Experimental and predicted $s$-state excitonic energies, together with the retrieved material parameters $E_g$, $r_0$, $\mu$, and $\kappa$, for WS$_2$, MoS$_2$, MoSe$_2$, and MoTe$_2$ monolayers in different dielectric environments, compared with the corresponding experimental and theoretical values reported in the literature.
}
\begin{ruledtabular}
\begin{tabular}{llccccc cccc}
Reference &Material &\multicolumn{5}{c}{Experimental and predicted energies (eV)} &\multicolumn{4}{c}{Retrieved parameters} \\
\cline{3-7}\cline{8-11}
\noalign{\vskip2pt}
&&$E_{1}$ &$E_{2}$ &$E_{3}$ &$E_{4}$ &$E_{5}$ &$E_g$ (eV) &$r_0$ (nm) &$\mu$ &$\kappa$ \\
\hline
\noalign{\vskip3pt}
Goryca et al.~\cite{NAT2019}& WS$_2$/hBN--hBN & & & & & & & & & \\
\quad -- Exp.& & 2.058\,\,\; & 2.197\,\,\; & 2.220\,\,\; &-- & --& -- & -- & -- & -- \\
\quad -- Theory& & 2.0585 & 2.1973 & 2.2200 & 2.2285 & --&2.238&3.40&0.175 &4.35\\
Theory~\cite{Molas2019}& & 2.0791 & 2.2028 & 2.2229 & 2.2297 & 2.2327 &-- &-- & --&-- \\
Retrieval~\cite{Nhat2025}& & 2.058\,\,\; & 2.197\,\,\; & 2.220\,\,\; & -- & -- & 2.239&3.40 &0.175 &4.10 \\
\textbf{This work}& & 2.058\,\,\; & 2.197\,\,\; & 2.220\,\,\; & 2.228\,\,\; & 2.231\,\,\; & 2.237 & 3.89 & 0.18 & 4.10 \\[3pt]

Chernikov et al.~\cite{chernikov2014}& WS$_2$/Vacuum--SiO$_ 2$ & & & & & & & & &\\
\quad -- Exp.& & 2.087\,\,\; & 2.251\,\,\; & 2.305\,\,\; & 2.344\,\,\; & 2.36& -- &-- &--  & -- \\
\quad -- Fitting& & 1.9908 & 2.2524 & 2.3208 & 2.3462 & 2.3617 &$2.41\pm 0.04$ &7.50  &0.16\,\,\;  &1.55\\
\textbf{This work}& & 2.087\,\,\; & 2.251\,\,\; & 2.305\,\,\; & 2.329\,\,\;  & 2.342\,\,\;  & 2.370 & 6.43 & 0.17\,\,\; & 1.55 \\[3pt]

Hsu et al.~\cite{Hsu2DMater2019}& WS$_2$/Vacuum--Sapphire & & & & & & & & & \\
\quad -- Exp.& & 2.018\,\,\; & 2.198\,\,\; & 2.246\,\,\; &-- &-- & -- & -- & -- & -- \\
\quad -- Fitting& & --& -- & --&-- &-- & 2.30 & 4.90 & 0.16\,\,\;  & 2.07 \\
\textbf{This work}& & 2.018\,\,\; & 2.198\,\,\; & 2.246\,\,\; & 2.266 \,\,\;& 2.275\,\,\; & 2.295 & 5.15 & 0.18\,\,\; & 2.07 \\[3pt]

Goryca et al.~\cite{NAT2019}& MoS$_2$/hBN--hBN & & & & & & & & & \\
\quad -- Exp.& & 1.938\,\,\; & 2.108\,\,\; & 2.137\,\,\; & --& --& -- & --& --  & -- \\
\quad -- Theory& & 1.9369 & 2.1056 & 2.1359 & 2.1458 &-- &2.161 &3.40  &0.27\,\,\;&4.40 \\
Theory~\cite{Molas2019}& & 1.9449 & 2.1104 & 2.1390 & 2.1488 & 2.1532 & --&-- & --& --\\
Retrieval~\cite{Nhat2025} & & 1.938\,\,\; & 2.108\,\,\;  & 2.137\,\,\; &--  & -- & 2.160 & 3.37 & 0.275 & 4.43 \\
\textbf{This work}& & 1.938\,\,\; & 2.108\,\,\;  & 2.137\,\,\; & 2.147 \,\,\; & 2.151 \,\,\; & 2.159 & 3.37 & 0.28\,\,\; & 4.48\\[3pt]

Goryca {\it{et al.}}~\cite{NAT2019}& MoSe$_2$/hBN--hBN & & & & & & & & & \\
\quad -- Exp.& & 1.643\,\,\; & 1.813\,\,\; & 1.847\,\,\; & --&-- &  --& --& -- &  --\\
\quad -- Theory& & 1.6428 & 1.8143 & 1.8469 & 1.8588 &-- &1.874& 3.90 &0.35\,\,\; &4.40 \\
 Theory~\cite{Molas2019}& & 1.6483 & 1.8151 & 1.8475 & 1.8590 & 1.8643 &-- &-- & --& --\\
 Retrieval~\cite{Nhat2025} & & 1.643\,\,\; & 1.813\,\,\; & 1.847\,\,\; & -- & -- & 1.875 & 4.02 & 0.345 & 4.22 \\
\textbf{This work}& & 1.643\,\,\; & 1.813\,\,\; & 1.847\,\,\; & 1.859\,\,\;  & 1.865\,\,\;  & 1.875 & 4.33 & 0.355 & 4.24 \\[3pt]

Liu et al.~\cite{Liu2021}& MoSe$_2$/hBN--hBN & & & & & & & & & \\
\quad -- Exp.& & 1.6408 & 1.7904 & 1.8141 & --& --& --&-- &-- &-- \\
\quad -- Fitting &-- & --&-- &-- &-- &-- & --& --& 0.40\,\,\;  &-- \\
\textbf{This work}& & 1.6408 & 1.7904 & 1.8141 & 1.8220 & 1.8255 & 1.832 & 3.30 & 0.41\,\,\;  & 6.23 \\[3pt]

Biswas et al.~\cite{Biswas2023}& MoTe$_2$/hBN--hBN & & & & & & & & & \\
\quad -- Exp.& & 1.172\,\,\; & 1.290\,\,\; & 1.315\,\,\; &-- & --&-- & --& --&-- \\
Retrievel~\cite{NAT2019}& & 1.175\,\,\; & 1.299\,\,\; & --\,\,\; & -- \,\,\; & -- \,\,\; & 1.352 & $6.4\pm 0.3$\,\,\; & $0.36\pm 0.04$\,\,\;   & 4.4 \\
\textbf{This work}& & 1.172\,\,\; & 1.290\,\,\; & 1.315\,\,\; & 1.324 \,\,\; & 1.329 \,\,\; & 1.337 & 6.70 & 0.32\,\,\;   & 4.5 \\

\end{tabular}
\end{ruledtabular}
\end{table*}

The hBN-encapsulated WSe$_2$ monolayer provides the most comprehensive benchmark because five independent experimental datasets are available. Among them, the measurements of Stier \textit{et al.}~\cite{Stier2018} and Liu \textit{et al.}~\cite{Liu2019} constitute the most controlled tests, since both optical and magneto-optical spectra are available for retrieving all material parameters independently. In both cases, the present analytical retrieval reproduces the quasiparticle bandgap within only a few meV of the RK analysis, while the retrieved screening lengths, reduced exciton masses, and dielectric constants remain fully consistent with the expected hBN environment. For the dataset of Liu \textit{et al.}~\cite{Liu2019}, the dielectric constant inferred from the experimentally retrieved reduced mass is closer to the accepted value for hBN than that adopted in the original RK fitting. Moreover, the retrieved parameters agree closely with those obtained previously from our analytical retrieval based directly on the Rytova--Keldysh potential~\cite{Nhat2025}, demonstrating that the modified Kratzer model faithfully reproduces the material parameters of hBN-encapsulated WSe$_2$ while extending the applicability of analytical retrieval beyond the restricted screening regime of the RK approximation.

Further validation is provided by the datasets of Takahashi \textit{et al.}~\cite{takahashi2024}, Chen \textit{et al.}~\cite{Chennano2019}, and Molas \textit{et al.}~\cite{Molas2019}. For the Takahashi experiment, the retrieved quasiparticle bandgap differs from the RK fitting by only 3~meV, while the screening length lies within the uncertainty of the reported RK value. Furthermore, the retrieved parameters are nearly identical to those obtained previously from our RK-based analytical retrieval, providing one of the strongest validations of the present inversion scheme. For the Chen dataset, both the present work and our previous RK retrieval consistently predict a screening length slightly larger than that adopted in the original RK fitting, suggesting that the difference is more likely associated with the experimental dataset than with the analytical approximation itself. The largest deviations are observed for the Molas dataset, where the experimental exciton energies are extracted by digitizing the published spectra rather than taken from tabulated values. Nevertheless, the retrieved parameters remain physically reasonable and are consistent with those obtained previously from the RK analytical retrieval. These results demonstrate that the proposed inversion remains robust against moderate uncertainties in the experimental input data.

The applicability of the present retrieval is further demonstrated for WSe$_2$ monolayers embedded in dielectric environments different from hBN. For the pioneering room-temperature measurements of He \textit{et al.}~\cite{HePRL2014}, the exciton energies were digitized from the published absorption spectra because neither tabulated energies nor RK fitting parameters are available. Although the retrieved bandgap is approximately 40~meV smaller than the experimentally determined continuum edge and the retrieved reduced exciton mass is somewhat larger than the commonly accepted value, the predicted higher excitonic states remain in reasonable agreement with experiment, indicating that the analytical retrieval still captures the essential spectral characteristics. Considerably better agreement is obtained for the more recent measurements of Hsu \textit{et al.}~\cite{Hsu2DMater2019}. For the PDMS-supported sample, the retrieved quasiparticle bandgap coincides with the RK result, while the screening length differs by only about $7\%$. For the sapphire-supported sample, the retrieved bandgap again agrees excellently with the RK analysis, whereas the remaining differences in the screening length and reduced exciton mass can be attributed primarily to the uncertainty associated with digitizing the published spectra. These results demonstrate that the present analytical retrieval remains reliable over a broad range of dielectric environments extending well beyond hBN encapsulation.

The applicability of the analytical retrieval is further confirmed by the WS$_2$ and Mo-based TMDC monolayers summarized in Table~\ref{tab2}. For hBN-encapsulated WS$_2$, MoS$_2$, and MoSe$_2$, the retrieved material parameters agree closely with both the published RK calculations and our previous RK-based analytical retrieval. The screening lengths obtained from the modified Kratzer model are generally slightly larger than those derived directly from the RK potential, reflecting the approximate nature of the analytical potential, whereas the quasiparticle bandgaps and reduced exciton masses remain essentially identical. The vacuum-supported WS$_2$ samples investigated by Chernikov \textit{et al.}~\cite{chernikov2014} and Hsu \textit{et al.}~\cite{Hsu2DMater2019} further demonstrate that the present analytical retrieval remains applicable in dielectric environments where our previous RK analytical retrieval is no longer valid. An independent validation is also provided by the second MoSe$_2$ dataset of Liu \textit{et al.}~\cite{Liu2021}, while the successful retrieval for hBN-encapsulated MoTe$_2$ extends the applicability of the method to the complete family of representative semiconducting TMDC monolayers.

In summary, the results presented in Tables~\ref{tab1} and \ref{tab2} demonstrate that the proposed analytical retrieval provides a robust approach for determining the fundamental material parameters of semiconducting TMDC monolayers from optical and magneto-optical spectroscopy. The two-stage retrieval procedure allows the quasiparticle bandgap and screening length to be determined from the zero-field exciton spectrum, while the reduced exciton mass is independently extracted from the magnetic-field dependence of the exciton energies whenever magneto-optical data are available. Across a broad range of TMDC materials and dielectric environments, the retrieved parameters are in good agreement with independent experimental measurements and previous Rytova--Keldysh analyses. The remaining discrepancies are mainly associated with digitized experimental data or the absence of independent benchmark parameters. Having established the reliability of the retrieved material parameters, we next examine their predictive capability for other excitonic properties.

\subsection{Prediction of derived excitonic properties}
\label{subsec:3c}

\begin{figure*}[htbp!]
\center
\includegraphics[width=2 \columnwidth]{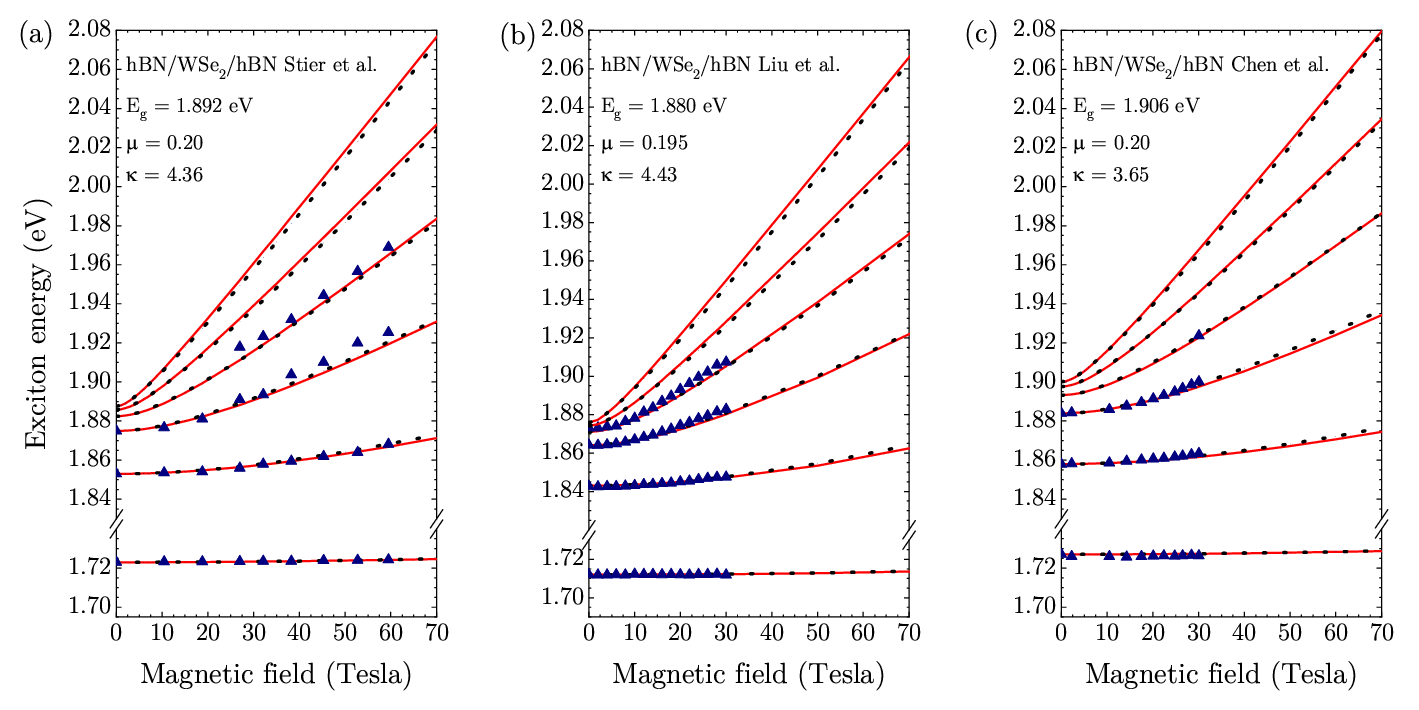}
\caption{Magnetoexciton energy spectra of monolayer WSe$_2$. Analytical results obtained from Eq.~(\ref{equ7}) are shown by red solid lines, numerical solutions by black dotted lines, and experimental data from Refs.~\cite{Stier2018, Liu2019, Chennano2019} by navy symbols.}
\label{fig3}
\end{figure*}

\begin{figure*}[htbp!]
\center
\includegraphics[width=2 \columnwidth]{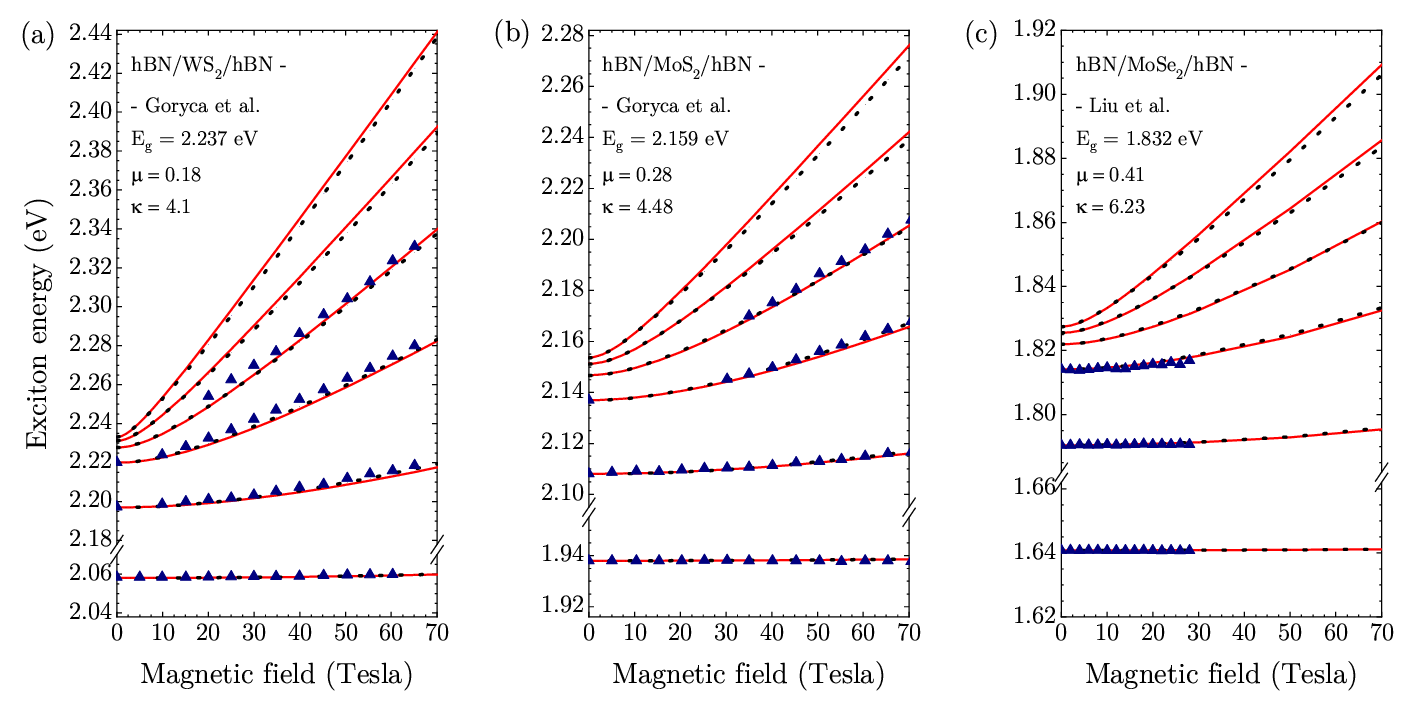}
\caption{Magnetoexciton energy spectra of monolayer WS$_2$, MoS$_2$, and MoSe$_2$. Analytical results obtained from Eq.~(\ref{equ7}) are shown by red solid lines, numerical solutions by black dotted lines, and experimental data by navy symbols. The experimental results for WS$_2$ and MoS$_2$ are taken from Ref.~\cite{NAT2019}, while those for MoSe$_2$ are from Ref.~\cite{Liu2021}.}
\label{fig4}
\end{figure*}

Once the complete set of material parameters has been retrieved, the present analytical framework directly predicts a variety of derived excitonic properties. These include the root-meam-square (rms) exciton radii, diamagnetic coefficients, and the complete magnetic-field dependence of the exciton spectrum. Since none of these quantities enters the retrieval procedure described in Subsection~\ref{subsec:3a}, they provide independent tests of the analytical model. Table~\ref{tab3} summarizes the predicted diamagnetic coefficients and rms exciton radii together with the available experimental and theoretical results, while Figs.~\ref{fig3} and~\ref{fig4} compare the analytical magnetoexciton spectra with both experimental measurements and direct numerical solutions of the Schr\"odinger equation. The excellent agreement demonstrates that the analytical interpolation formula accurately reproduces the magnetic response of excitons over the experimentally accessible magnetic-field range.

% Table III
\begin{table*}[htbp]
\caption{\label{tab3}
Diamagnetic coefficients $\sigma_n$ and exciton radii $r_n$ for the four lowest $s$ excitonic states, compared with the corresponding experimental and theoretical values reported in the literature. The diamagnetic coefficients are given in $\mu$eV/T$^2$, and the exciton radii in nm.}
\begin{ruledtabular}
\begin{tabular}{l cc cc cc cc}
Reference/Material & $\sigma_1$ & $r_1$ &$\sigma_2$ & $r_2$ &$\sigma_3$ & $r_3$ &$\sigma_4$ & $r_4$  \\
\hline
\noalign{\vskip3pt}
Stier et al.~\cite{Stier2018} & \multicolumn{8}{c}{WSe$_2$/hBN--hBN}\\
\quad -- Exp.     & $0.31\pm0.02$ & $1.7\pm0.1$ & $4.6\pm0.2$ & $6.6\pm0.4$ & $22\pm2$ & $14.3\pm1.5$ &-- & -- \\
\quad -- Theory & 0.30 & 1.67 & 5.12 & 6.96 & 26.86 & 15.80 & &  \\
Retrieval~\cite{Nhat2025}  & 0.34 & 1.72 & 5.67 & 7.01 & 29.32 & 15.91 & 94.16 & 28.52  \\
\textbf{This work}  & 0.35 & 1.76 & 5.42 & 6.99 & 27.96 & 15.87 & 89.67 & 28.42  \\[3pt]

Liu et al.~\cite{Liu2019}  & \multicolumn{8}{c}{WSe$_2$/hBN--hBN} \\
\quad -- Exp.    & $0.24\pm0.01$ & $1.6\pm0.4$ & $6.4\pm0.2$ & $8.24\pm0.13$ & $27.3\pm1.3$ & $17.0\pm0.4$ & $73.7\pm3.0$ & $27.8\pm0.7$  \\
\quad -- Theory  & 0.31 & 1.68 & 4.86 & 6.66 & 24.20 & 14.86 & 76.30 & 26.37  \\
Retrieval~\cite{Nhat2025}   & 0.34 & 1.69 & 5.63 & 6.98 & 29.39 & 15.94 & 94.88 & 28.64  \\
\textbf{This work}  & 0.35 & 1.75 & 5.64 & 7.07 & 29.51 & 16.18 & 95.35 & 29.08  \\[3pt]

Chen et al.~\cite{Chennano2019} & \multicolumn{8}{c}{WSe$_2$/hBN--hBN} \\
\quad -- Exp.     & 0.50 & 2.20 & 5.80 & 7.60 & 17.60 & 13.30 & &  \\
\quad -- Theory  & 0.25 & 1.60 & 4.18 & 6.50 & 21.60 & 14.70 & &  \\
Retrieval~\cite{Nhat2025}  & 0.35 & 1.73 & 4.92 & 6.52 & 23.13 & 14.13 & 70.39 & 24.66  \\
\textbf{This work}  & 0.38 & 1.86 & 4.87 & 6.66 & 23.14 & 14.51 & 71.02 & 25.42  \\[3pt]

Goryca et al.~\cite{NAT2019}& \multicolumn{8}{c}{WS$_2$/hBN--hBN} \\
\quad -- Theory  & & 1.80 & & & & & & \\
Retrieval~\cite{Nhat2025}  & 0.37 & 1.71 & 6.26 & 7.06 & 33.41 & 16.30 & 109.30 & 29.49  \\
\textbf{This work}  & 0.38 & 1.77 & 6.20 & 7.13 & 32.37 & 16.28 & 104.44 & 29.24  \\[3pt]

Goryca et al.~\cite{NAT2019}& \multicolumn{8}{c}{MoS$_2$/hBN--hBN} \\
\quad -- Theory  & 0.12 & 1.20 &-- & --&-- &-- & --&  --\\
Retrieval~\cite{Nhat2025}  & 0.12 & 1.22 & 2.05 & 5.06 & 10.68 & 11.55 & 34.42 & 20.75  \\
\textbf{This work}  & 0.13 & 1.30 & 2.06 & 5.13 & 10.61 & 11.62 & 33.94 & 20.79  \\[3pt]

Goryca et al.~\cite{NAT2019}& \multicolumn{8}{c}{MoSe$_2$/hBN--hBN} \\
\quad -- Theory  & 0.07 & 1.10 & & & & & &  \\
Retrieval~\cite{Nhat2025}   & 0.08 & 1.12 & 1.18 & 4.30 & 5.71 & 9.47 & 17.73 & 16.68  \\
\textbf{This work}  & 0.09 & 1.23 & 1.19 & 4.38 & 5.62 & 9.52 & 17.20 & 16.66  \\[3pt]

Liu et al.~\cite{Liu2021}& \multicolumn{8}{c}{MoSe$_2$/hBN--hBN} \\
\quad -- Theory  & & 1.10 & & 3.20 & & 8.10 & &  \\
\textbf{This work}  & 0.07 & 1.12 & 1.14 & 4.61 & 6.06 & 10.63 & 19.77 & 19.20  \\[3pt]

Biswas et al.~\cite{Biswas2023} & \multicolumn{8}{c}{MoTe$_2$/hBN--hBN} \\
\quad -- Theory~\cite{NAT2019}  & 0.10 & 1.30 & 1.40 & & & & &  \\
\textbf{This work}  & 0.14 & 1.50 & 1.62 & 5.15 & 7.42 & 11.03 & 22.35 & 19.13  \\

\end{tabular}
\end{ruledtabular}
\end{table*}

The hBN-encapsulated WSe$_2$ monolayer provides the most stringent benchmark because several independent magneto-optical experiments report both diamagnetic coefficients and exciton radii. The measurements of Stier \textit{et al.}~\cite{Stier2018} and Liu \textit{et al.}~\cite{Liu2019} constitute the most comprehensive datasets, allowing direct comparison up to the $4s$ exciton. As shown in Table~\ref{tab3}, the present analytical model reproduces the measured exciton radii and diamagnetic coefficients with good accuracy throughout the Rydberg series. The largest deviation occurs for the $4s$ diamagnetic coefficient reported by Liu \textit{et al.}, which is expected because the magnetic response of such a highly excited state is particularly sensitive to both the interaction potential and the magnetic-field fitting procedure. Overall, the predicted magnetic properties remain in excellent agreement with our previous RK-based analytical retrieval~\cite{Nhat2025}, demonstrating that the modified Kratzer model preserves the predictive capability of the RK approach while providing a much simpler analytical framework. The present model also predicts the diamagnetic coefficients and exciton radii of higher Rydberg states that have not yet been measured experimentally.

Further validation is provided by the measurements of Chen \textit{et al.}~\cite{Chennano2019}. Unlike the datasets of Stier \textit{et al.}~\cite{Stier2018} and Liu \textit{et al.}~\cite{Liu2019}, the exciton radii were estimated by assuming a reduced exciton mass of $\mu=0.22$. Nevertheless, the present calculations reproduce both the measured diamagnetic coefficients and the estimated exciton radii with good accuracy, remaining consistent with the RK calculations of Ref.~\cite{Chennano2019} and our previous RK-based analytical retrieval~\cite{Nhat2025}. The analytical framework further predicts the magnetic properties of the $4s$ exciton, for which neither experimental nor theoretical results are currently available. These results further demonstrate that the modified Kratzer model reliably predicts both the spatial and magnetic properties of excitons.

For WS$_2$, two complementary experiments provide independent validation of the present predictions. The high-field magneto-optical measurements of Goryca \textit{et al.}~\cite{NAT2019} confirm the predicted ground-state exciton radius and remain consistent with our previous RK-based analytical retrieval, while the present framework further predicts the complete series of exciton radii and diamagnetic coefficients up to the $4s$ state. Complementary validation is provided by the magneto-optical measurements of Chernikov \textit{et al.}~\cite{chernikov2014}. As shown in Fig.~\ref{fig4}, the analytical magnetoexciton spectra closely reproduce the experimental magnetic-field dependence and are almost indistinguishable from the direct numerical solutions of the Schr\"odinger equation. Together, these results demonstrate that the analytical framework accurately predicts both the magnetic properties and the field-dependent evolution of excitons in monolayer WS$_2$.

The predictive capability of the analytical framework is further confirmed for the Mo-based TMDC monolayers. For hBN-encapsulated MoS$_2$ and MoSe$_2$, the magneto-optical measurements of Goryca \textit{et al.}~\cite{NAT2019} validate the predicted ground-state exciton radii, while the remaining higher-state quantities remain consistent with our previous RK-based analytical retrieval. Additional validation is provided by the magneto-photoluminescence measurements of Liu \textit{et al.}~\cite{Liu2021}, whose extracted $2s$ and $3s$ exciton radii in monolayer MoSe$_2$ confirm the predicted increase of the exciton size with principal quantum number, although the calculated values are systematically somewhat larger. Furthermore, Fig.~\ref{fig4} shows that the analytical magnetoexciton spectra closely reproduce both the experimental observations and the corresponding numerical solutions of the Schr\"odinger equation. Together, these results demonstrate the robustness of the present analytical framework across the Mo-based TMDC family.

The final example is provided by hBN-encapsulated MoTe$_2$, where the zero-field exciton spectrum reported by Biswas \textit{et al.}~\cite{Biswas2023} is combined with the available magneto-optical data of Goryca \textit{et al.}~\cite{NAT2019}. The predicted values of $\sigma_{1s}$, $\sigma_{2s}$, and $r_{1s}$ agree reasonably well with the available measurements, while the corresponding properties of the higher Rydberg states constitute genuine predictions awaiting future experimental verification. This example demonstrates that the analytical framework remains predictive even when the available magneto-optical information is incomplete.

In total, the results presented in Table~\ref{tab3} and Figs.~\ref{fig3} and \ref{fig4} provide a comprehensive validation of the predictive capability of the proposed analytical framework. Since neither the diamagnetic coefficients nor the exciton radii enters the retrieval procedure, their good agreement with available experimental measurements and previous RK-based calculations constitutes an independent verification of the retrieved material parameters. Furthermore, the excellent agreement between the analytical magnetoexciton spectra and the corresponding numerical solutions of the Schr\"odinger equation demonstrates that the analytical interpolation formula accurately describes the magnetic-field evolution of excitons over the experimentally accessible field range. Having established the predictive capability of the modified Kratzer model, we next examine its consistency with the Rytova--Keldysh description.

% Figure 3

%Figure4

\subsection{Validation of the modified Kratzer model}

Having demonstrated the accuracy of the retrieved material parameters and the predicted excitonic properties, we now examine the consistency between the modified Kratzer and Rytova--Keldysh (RK) descriptions. This comparison provides a direct assessment of the applicability of the modified Kratzer model as an analytical representation of the screened electron--hole interaction in monolayer TMDCs.

%Figure 5
\begin{figure*}[htbp]
\begin{center}
\includegraphics[width=1.7 \columnwidth]{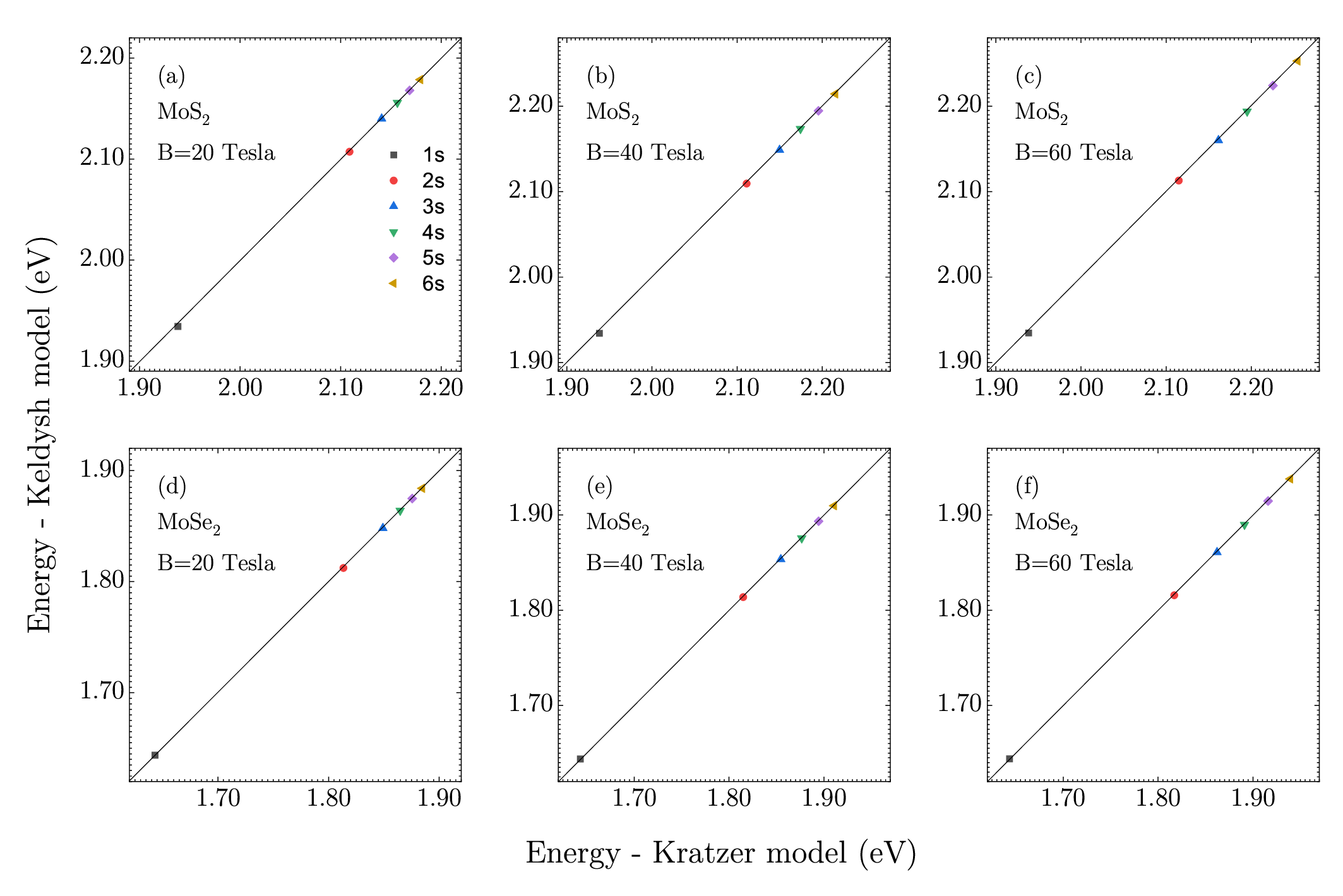}
\caption{Comparison of magnetoexciton energy levels obtained using the Kratzer model (horizontal axis) and the Rytova--Keldysh model (vertical axis) for MoS$_2$ and MoSe$_2$, using the material parameters in Tables~\ref{tab2}. Panels (a--c) correspond to MoS$_2$ at magnetic fields $B = 20$, 40, and 60~T, respectively, while panels (d--f) show the corresponding results for MoSe$_2$ at the same magnetic fields.}
\label{fig5}
\end{center}
\end{figure*}

Figure~\ref{fig5} compares the magnetoexciton energies calculated using the modified Kratzer and RK potentials. An excellent one-to-one correspondence is obtained over the entire range of investigated magnetic fields and excitonic states, indicating that the modified Kratzer model accurately reproduces the RK magnetoexciton spectrum. The agreement extends well beyond the low-lying states used in the retrieval procedure, demonstrating that the analytical framework reliably describes both the ground and higher Rydberg excitons.

A further comparison is provided by the retrieved material parameters summarized in Tables~\ref{tab1} and \ref{tab2}. For nearly all investigated TMDC monolayers and dielectric environments, the quasiparticle bandgap $E_g$, reduced exciton mass $\mu$, and surrounding dielectric constant $\kappa$ obtained from the modified Kratzer model agree closely with previous RK analyses and independent experimental determinations. The most noticeable difference appears in the retrieved screening length $r_0$, which is generally reproduced more accurately by the RK-based retrieval. This discrepancy originates from the universal approximation $g^2=0.205$ adopted in the modified Kratzer model, which allows a simple analytical inversion over a broad range of material parameters at the expense of a modest reduction in the accuracy of $r_0$.

Nevertheless, the influence of this approximation on experimentally observable quantities is very limited. As demonstrated in Subsecs.~\ref{subsec:3a} and \ref{subsec:3b}, the retrieved exciton energies, magnetoexciton spectra, diamagnetic coefficients, and exciton radii remain in excellent agreement with both experiments and RK calculations. The modified Kratzer model therefore provides an effective compromise between analytical simplicity, broad applicability, and predictive accuracy.

The remaining discrepancy in the screening length also suggests a natural direction for future work. An important objective will be to develop a generalized analytical retrieval directly from the RK potential that preserves the broad applicability achieved by the present approach while retaining the higher quantitative accuracy of the RK description.

%========================================================
\section{Conclusion}
\label{sec:4}
%========================================================

In this work, we have developed a comprehensive analytical framework to directly retrieve fundamental material parameters of monolayer transition-metal dichalcogenides from optical and magneto-optical spectroscopy. Based on the exactly solvable modified Kratzer model, explicit inversion formulas have been derived to determine material parameters such as the quasiparticle bandgap, screening length, reduced exciton mass, and the surrounding dielectric constant. 

The proposed retrieval procedure naturally consists of two complementary stages. In the first stage, the quasiparticle bandgap and screening length are uniquely determined from the zero-field energies of the three lowest excitonic states. In the second stage, the reduced exciton mass is independently extracted from the magnetic-field dependence of the exciton energies using an analytical magnetoexciton expression, from which the surrounding dielectric constant is subsequently obtained. 

This two-stage retrieval procedure has been successfully applied to a broad range of experimental datasets to determine the fundamental material parameters of representative TMDC monolayers. Once the material parameters have been retrieved, the analytical framework predicts the remaining excitonic properties, including the diamagnetic coefficients, exciton radii, and complete magnetoexciton spectra, without introducing additional fitting parameters. 

The excellent agreement between the predicted quantities and available experimental measurements, together with our previous calculations based on the Rytova--Keldysh model for representative TMDC monolayers embedded in various dielectric environments, demonstrates both the accuracy and the broad applicability of the proposed approach. Moreover, the close agreement between the modified Kratzer and Rytova--Keldysh models establishes a quantitative correspondence between the two descriptions over the experimentally relevant parameter range. 

Beyond the materials investigated in the present work, the proposed framework provides a practical analytical tool for the rapid characterization of two-dimensional semiconductors using optical and magneto-optical spectroscopy. Owing to its computational efficiency, physical transparency, and broad applicability across diverse dielectric environments, the method should prove useful for analyzing future excitonic experiments and may also serve as a foundation for developing more general analytical retrieval schemes based directly on the Rytova--Keldysh interaction.
\section*{Acknowledgments} This research is funded by Vietnam Ministry of Education and Training, Grant number B2025-SPS-03 and
carried out by the high-performance cluster at Ho Chi Minh City University of Education, Vietnam. 

Contribution: V.-H.Le and D.-N.Ly conceptualized the work and developed the methodology; D.-N.Ly and V.-H.Le performed analytical formulation; D.-N.Ly carried out numerical calculations; T.-S.N. and N.-T.D.H. validated the data; D.-N.Ly and V.-H.Le wrote the original draft; N.-T.D.H. acquired the funding. All authors discussed the results and contributed to the review, editing, and finalization of the manuscript.

\section*{Data availability}
The data that support the findings of this article are not publicly available. The data are available from the authors upon reasonable request.

\appendix
\section{Numerical solution of the magnetoexciton Schrödinger equation in the modified Kratzer model}\label{AppA}

To assess the accuracy of the analytical magnetoexciton expression derived in Subsec.~\ref{subsec:2c}, we numerically solve the Schr{\"o}dinger equation for the modified Kratzer model in the presence of a perpendicular magnetic field. The resulting numerical energies serve as benchmark solutions for the comparisons presented in Figs.~\ref{fig3} and \ref{fig4}. This Appendix summarizes the numerical procedure.

We consider the radial Schr\"odinger equation 
\begin{equation}\label{a1}
\left\{\frac{\partial^{2}}{\partial r^{2}}+ \frac{1}{r}\frac{\partial}{\partial r}- \frac{\xi^{2}}{r^{2}}
+ \frac{2}{\kappa r} - \frac{\gamma^{2}}{4}r^{2} + \varepsilon\right\} \mathcal{R}(r) = 0 , 
\end{equation}
where $\mathcal{R}(r)$ is the radial wave function. The equation is written in effective atomic units, where $\gamma$, $\varepsilon$, $r = \sqrt{x^2 + y^2}$ are the magnetic-field strength, exciton binding energy, and exciton radius respectively in units of $B_{0}^{*}=\mu^{2}B_0$, $\text{Ry}^{*}=\mu \text{Ry}$, and $a_{0}^{*}=a_0/\mu$, scaling by the dimensionless reduced exciton  mass $\mu$ (in units of the electron mass $m_e$). Here, the standard notations are used for the Bohr radius $a_0 = 4\pi\varepsilon_0 \hbar^{2}/(m_e e^{2})$, the Rydberg energy $\text{Ry} = m_e e^{4}/(32\pi^{2}\varepsilon_{0}^{2}\hbar^{2})$, and the magnetic-field unit $B_0 =2 m_e \text{Ry}/(e \hbar)$. 

Equation~\eqref{a1} is established in the frame of the modified Kratzer model \cite{Molas2019}, where the electron--hole interaction in 2D semiconductors, conventionally described by the Rytova-Keldysh potential~\cite{Rytova1967,keldysh1979}, is approximated by the Coulomb potential $-{1}/{\kappa r}$ added a centrifugal term ${\xi^2}/{r^2} $. This repulsive part of the interaction potential is reponsible for the screening effect, so we call $\xi$ the effective screening parameter. The Coulomb potential is amplified by the dielectric konstant $\kappa$ of the surrounding medium.

In the absence of a magnetic field ($\gamma=0$), Eq.~\eqref{a1} reduces to an equation for the 2D hydrogen-like problem in which the centrifugal term is added by the repulsive contribution $\xi^{2}/r^{2}$. The corresponding eigenvalues for $s$ states are
\begin{equation}\label{a2}
\varepsilon_{n}= -\frac{1}{\kappa^{2}\left( n  - \tfrac{1}{2} + \xi \right)^{2}},
\end{equation}
with $n=1,2,\ldots$ is the principal quantum number. By changing from unit of effective Rydberg energy Ry$^*$ to Ry, the reduced exciton mass $\mu$ appears in the numerator of Eq.~\eqref{a2}, that we rewrite the energy expression as in Eq.~\eqref{eq1} in the main text, where we use the notation $\eta=\mu/\kappa^2$.
The normalized radial functions take the form
\begin{eqnarray}\label{a3}
\mathcal{R}_{n}(r)= \mathcal{A}_{n}\,\omega^{\xi+1} \,\mathrm{e}^{-\omega r/2}\,r^{\xi}\,
\mathcal{L}_{n}^{2\xi}(\omega r),
\end{eqnarray}
where $\mathcal{A}_{n} = \sqrt{ n!/(n+2\xi)!/(2n+2\xi+1) }$ is the normalization constant, $\mathcal{L}_{n}^{2\xi}(x)$ denotes the generalized Laguerre polynomials, and $\omega = 4/((2n+2\xi+1)\kappa)$.

For finite magnetic fields, the radial wave function is expanded in an orthonormal Laguerre basis,
\begin{eqnarray}\label{a4}
|R_{n}(r)\rangle
= \sum_{j=0}^{j_{\mathrm{max}}} C_{jn}\,|f_{j}(r)\rangle .
\end{eqnarray}
Here, $C_{jn}$ are the expansion coefficients and the basis functions are
\begin{eqnarray}\label{a5}
f_{j}(r)= \omega^{\xi + \tfrac{1}{2}}\sqrt{\frac{j!}{(j + 2\xi)!}}\,\mathrm{e}^{-\omega r/2}\,r^{\xi}\,
\mathcal{L}_{j}^{2\xi}(\omega r),
\end{eqnarray}
with an adjustable variational parameter $\omega$ used to optimize the convergence of the basis expansion, and the truncation index $j_{\mathrm{max}}$ in Eq.~\eqref{a4} chosen according to the desired numerical accuracy~\cite{nhat2021}.

The inner product in the radial Hilbert space is taken with the standard 2D polar-coordinate measure $r\,dr$, so that the overlap and Hamiltonian matrix elements involve the Jacobien $J=r$ rather than the identity. This accounts for the extra power of $r$ in the normalization and mean-square-radius expressions below. The normalization condition therefore becomes
\begin{eqnarray}\label{a6} \nonumber
&& \langle R_{n} | R_{n} \rangle
= \sum_{k=0}^{j_{\text{max}}} \sum_{j=0}^{j_{\text{max}}} C_{kn} C_{jn} 
\langle f_{k}(r)|\, J \,| f_{j}(r)\rangle \; \\
&& \qquad \qquad \qquad = \sum_{k=0}^{j_{\text{max}}} \sum_{j=0}^{j_{\text{max}}} 
C_{kn} J_{k j} C_{jn}= 1
\end{eqnarray}
with $J_{kj}=\langle f_{k}(r)|\,r\,|f_{j}(r)\rangle$.
The dimensionless mean-square radius is
\begin{eqnarray}\label{a7}
&& \langle R_{n}|r^2| R_{n} \rangle= \sum_{k=0}^{j_{\text{max}}} \sum_{j=0}^{j_{\text{max}}} 
C_{kn} {J_{k j}}^3 C_{jn},\;
\end{eqnarray}
where ${J_{k j}}^3=\langle f_{k}(r)|\,r^{3}\,|f_{j}(r)\rangle$.

At $\gamma=0$, the expansion coefficients reduce to the analytical form
\begin{eqnarray}\label{a8}
C_{jn} 
= \frac{2}{{\left( {2n + 2\xi  + 1} \right)\sqrt \kappa  }}\delta_{jn} ,
\end{eqnarray}
and the radial wave function Eq.~(\ref{a4}) reduces to Eq.~(\ref{a3}), from which the corresponding energy is recovered as given in Eq.~(\ref{a2}).

For finite fields, substitution of Eq.~(\ref{a4}) into Eq.~(\ref{a1}) and projecting onto the basis $\langle r\,f_{k}(r)\rvert$ yields the generalized eigenvalue problem
\begin{eqnarray}\label{a9}
\bigl(\mathbb{H}_0 + \mathbb{V}_{\mathrm{mag}} - {\varepsilon}\,\mathbb{R}\bigr)\,\mathbb{C} = 0,
\end{eqnarray}
where $\mathbb{C}$ denotes the column vector of expansion coefficients. The matrix $\mathbb{H}_0$ represents the nonmagnetic Hamiltonian, with elements
\begin{eqnarray}\label{a10} \nonumber
&&(H_0)_{kn}= -\frac{1}{\kappa}\,\delta_{k n} + \; \frac{\omega}{8}\sum_{i=0}^{1}\sqrt{\binom{n+i}{i}\,\binom{n+2\xi}{1-i}} \\ 
&&  \qquad \sum_{j=0}^{1}\sqrt{\binom{n+i-j+2\xi}{1-j}\,\binom{n+i}{j}}\,\delta_{k,\,n+i-j}. \qquad
\end{eqnarray}
The magnetic-field contribution is represented by the matrix $\mathbb{V}_{\mathrm{mag}}$, whose elements are
\begin{eqnarray}\label{a11}
&&(V_{\mathrm{mag}})_{k n}=\frac{3\gamma^{2}}{4\omega^{3}}\sum_{i=0}^{3}(-1)^{i}
\sqrt{\binom{n+i}{i}\,\binom{n+2\xi}{3-i}\,\binom{3}{i}}\sum_{j=0}^{3}\nonumber\\
&&(-1)^{j}\sqrt{\binom{n+i-j+2\xi}{3-j}\,\binom{n+i}{j}\,\binom{3}{j}}\,\delta_{k,\,n+i-j}. \qquad
\end{eqnarray}
The matrix $\mathbb{R}$ accounts for the Jacobian from the radial integration, with elements
\begin{eqnarray}\label{a12}
&&R_{k n}= \frac{1}{\omega}\sum_{i=0}^{1}(-1)^{i}\sqrt{\binom{n+i}{i}\,\binom{n+2\xi}{1-i}}\sum_{j=0}^{1} (-1)^{j}\nonumber\\
&& \qquad\times \sqrt{\binom{n+i-j+2\xi}{1-j}\,\binom{n+i}{j}}\,\delta_{k,\,n+i-j}. \qquad
\end{eqnarray}

Diagonalization of Eq.~(\ref{a9}) yields the magnetoexciton spectrum, used as benchmark results for validating the analytical magnetoexciton expression [Eq.~\ref{eq14}] in Figs.~\ref{fig3} and \ref{fig4}.

\bibliography{ref}% Produces the bibliography via BibTeX.

\end{document}